\documentclass[a4paper,10pt,twoside]{cpc-hepnp}
\usepackage{CJK,upgreek,fancyhdr}
\usepackage{multicol}
\usepackage{graphicx}
\usepackage{booktabs}
\usepackage{amssymb,bm,mathrsfs,bbm,amscd}
\usepackage[tbtags]{amsmath}
\usepackage{lastpage}
\usepackage{longtable}
\usepackage{changes}

\begin{document}
\begin{CJK*}{GB}{gbsn}
%\begin{CJK*}{GBK}{song}

%\fancyhead[c]{\small Chinese Physics C~~~Vol. xx, No. x (201x) xxxxxx}
%\fancyfoot[C]{\small 010201-\thepage}

%\footnotetext[0]{Received 2 July 2016}
\title{Systematic study of proton radioactivity of spherical proton emitters within various versions of proximity potential formalisms
\thanks{Supported by the National Natural Science Foundation of China (Grants No. 11205083 and No. 11505100), the Construct Program of the Key Discipline in Hunan Province, the Research Foundation of Education Bureau of Hunan Province, China (Grant No. 15A159), the Natural Science Foundation of Hunan Province, China (Grants No. 2015JJ3103 and No. 2015JJ2121), the Innovation Group of Nuclear and Particle Physics in USC, the Shandong Province Natural Science Foundation, China (Grant No. ZR2015AQ007), Hunan Provincial Innovation Foundation for Postgraduate (Grant No. CX2017B536), University of South China Science Foundation for Postgraduate (2018KYY027).}}

\author{%
   Jun-Gang Deng$^{1}$
\quad Xiao-Hua Li$^{1,3,4;1)}$\email{lixiaohuaphysics@126.com}
\quad Jiu-Long Chen$^{1}$
\quad Jun-Hao Cheng$^{1}$\\
\quad Xi-Jun Wu$^{2;2)}$\email{{wuxijun1980@yahoo.cn}}
}
\maketitle

\address{%
$^1$ School of Nuclear Science and Technology, University of South China, Hengyang 421001, China\\
$^2$ School of Math and Physics, University of South China, Hengyang 421001, China\\
$^3$ Cooperative Innovation Center for Nuclear Fuel Cycle Technology $\&$ Equipment, University of South China, Hengyang 421001, China\\
$^4$ Key Laboratory of Low Dimensional Quantum Structures and Quantum Control, Hunan Normal University, Changsha 410081, China\\
}

\begin{abstract}
In this work we present a systematic study of the proton radioactivity half-lives of spherical proton emitters within the Coulomb and proximity potential model. We investigate 28 different versions of the proximity potential formalisms developed for the description of proton radioactivity, $\mathcal{\alpha}$ decay and heavy particle radioactivity. It is found that 21 of them are not suitable to deal with the proton radioactivity, because the classical turning points $r_{\text{in}}$ cannot be obtained due to the fact that the depth of the total interaction potential between the emitted proton and the daughter nucleus is above the proton radioactivity energy. Among the other 7 versions of the proximity potential formalisms, it is Guo2013 which gives the lowest rms deviation in the description of the experimental half-lives of the known spherical proton emitters. We use this proximity potential formalism to predict the proton radioactivity half-lives of 13 spherical proton emitters, whose proton radioactivity is energetically allowed or observed but not yet quantified, within a factor of 3.71.
\end{abstract}

%\begin{keyword}
%proton radioactivity, half-life, proximity potential formalism, Geiger-Nuttall law
%\end{keyword}

\begin{pacs}
23.50.+z, 21.10.Tg
\end{pacs}

%\footnotetext[0]{\hspace*{-3mm}\raisebox{0.3ex}{$\scriptstyle\copyright$}2018
%Chinese Physical Society and the Institute of High Energy Physics
%of the Chinese Academy of Sciences and the Institute
%of Modern Physics of the Chinese Academy of Sciences and IOP Publishing Ltd}%

\begin{multicols}{2}

\section{Introduction}
The study of exotic nuclei, one of the hot topics in contemporary nuclear physics, led to the discovery of a new decay mode--proton radioactivity. In 1970, Jackson \textit{et al.} firstly detected the proton radioactivity from the isomeric state of $^{53}$Co to the ground state of $^{52}$Fe in experiment \cite{JACKSON1970281}. With the development of experimental facility, the proton radioactivity of nuclei from ground state or low-lying isomer state has been observed in the mass number region $A=$110--180 \cite{SONZOGNI20021}. Proton radioactivity, a typical decay mode of odd-$Z$ nuclei beyond the proton drip line, limits the creation of more proton-rich nuclei in the proton side of $\mathcal{\beta}$-stability line. Moreover, the study of proton radioactivity can obtain some important information of nuclear structure such as the shell structure, the coupling between bound and unbound nuclear states \cite{KARNY200852} and so on. In addition, proton radioactivity half-life is sensitive to proton radioactivity energy and angular momentum carried out by emitted proton \cite{SONZOGNI20021}. Therefore, measurements of the half-life and decay energy of proton radioactivity help to determine the angular momentum taken away by the emitted proton and to characterize its wave function inside the nucleus \cite{PhysRevC.69.054311,L. S. Ferreira,PhysRevC.79.054330,1674-1137-34-2-005,0256-307X-26-7-072301}.

The proton radioactivity can be dealt with the Wentzel-Kramers-Brillouin (WKB) method because this process can be treated as a simple quantum tunneling effect through a potential barrier in the same way as $\mathcal{\alpha}$ decay. There are many methods to investigate the proton radioactivity such as the density-dependent M3Y effective interaction \cite{BHATTACHARYA2007263,0256-307X-27-7-072301,PhysRevC.72.051601}, the JLM interaction \cite{BHATTACHARYA2007263}, the unified fission model \cite{PhysRevC.71.014603,1674-1137-34-2-005}, the generalized liquid drop model (GLDM) \cite{PhysRevC.79.054330}, the cluster model \cite{0256-307X-26-7-072301}, the deformed density-dependent model \cite{Qian2016,0256-307X-27-11-112301}, the Gamow-like model \cite{Zdeb2016}, the Coulomb and proximity potential model for deformed nuclei (CPPMDN) \cite{PhysRevC.96.034619}, the covariant density functional (CDF) theory \cite{PhysRevC.90.054326}, the analytic formula \cite{1674-1137-42-1-014104}, the distorted-wave Born approximation (DWBA) \cite{PhysRevC.56.1762}, the two-potential approach (TPA) \cite{PhysRevC.56.1762}, the quasiclassical method \cite{PhysRevC.56.1762} and so on \cite{PhysRevC.45.1688,PhysRevLett.81.538,PhysRevC.59.R589}.

The proximity potential was put forward by Blocki \textit{et al.} to deal with heavy ion reaction \cite{BLOCKI1977427}. As a nucleus-nucleus interaction potential, it is based on the proximity force theorem \cite{BLOCKI1977427,BLOCKI198153}, which is described as the product of a factor depending on the mean curvature of the interaction surface and an universal function (depending on the separation distance) and is independent of the masses of colliding nuclei \cite{PhysRevC.81.044615}. It has the advantages of simple and accurate formalism. In order to overcome the shortcomings of the original version of the proximity potential (Prox.1977) \cite{BLOCKI1977427}, various improvements and modifications have been proposed. Those developments included either a better form of the surface energy coefficients \cite{MYERS19661,MOLLER1976502,PhysRevC.20.992,MOLLER1981117,0305-4616-10-8-011,MOLLER1988213,MOLLER1995185,PhysRevC.67.044316}, or introduced an improved universal function or another nuclear radius parameterization \cite{BLOCKI198153,PhysRevC.62.044610,0256-307X-27-11-112402,DUTT2011,BASS1973139,BASS197445,PhysRevLett.39.265,0954-3899-20-9-004,CHRISTENSEN197619,WINTHER1995203,NGO1980140,DENISOV2002315,PhysRevC.81.044615,GUO201354}.

Recently, comparative studies of various proximity potential formalisms for describing $\mathcal{\alpha}$ decay and heavy particle radioactivity have been performed by Yao \textit{et al.} \cite{Yao2015}, Ghodsi \textit{et al.} \cite{PhysRevC.93.024612} and Santhosh \textit{et al.} \cite{Santhosh2017}, respectively. In addition, the proximity potential formalisms were used to study the heavy ion fusion \cite{SANTHOSH200935,SANTHOSH2014191} and ternary fission \cite{0954-3899-41-10-105108,PhysRevC.91.044603}. In our previous works \cite{PhysRevC.97.044322,1674-1137-42-4-044102}, we investigated the $\mathcal{\alpha}$ decay within the proximity potential 1977 formalism \cite{BLOCKI1977427} and generalized proximity potential 1977 formalism (gp77, Prox.81) \cite{BLOCKI198153,PhysRevC.87.064611}. In these works, the nuclear potential can be obtained easily and the calculations can well conform to the experimental data by using the proximity potential formalism. Therefore, it has been demonstrated that the proximity potential formalism is a simple and convenient approach applicable in different domains of nuclear physics. It is interesting and important to show its capacities and performance to describe various decay modes. Therefore, the main objective of this work is to investigate the applicability of various proximity potentials to proton radioactivity of spherical proton emitters. The calculations indicate that the proximity potential Guo2013 formalism gives the lowest rms deviation in the description of the experimental half-lives of the known spherical protons emitters. Using this proximity potential formalism we predict the proton radioactivity half-lives of 13 spherical proton emitters, whose proton radioactivity is energetically allowed or observed, but has not been quantified yet. 

This article is organized as follows. In Section 2, the theoretical frameworks of proton radioactivity half-life and proximity potential formalism are briefly described. The detailed calculations and discussion are presented in Section 3. Finally, a summary is given in Section 4.

\section{Theoretical framework}
\label{section 2}
\subsection{The proton radioactivity half-life}
\label{section 2.1}
The proton radioactivity half-life is related to the decay constant $\mathcal{\lambda}$ as 
\begin{equation}
\
T_{1/2}=\frac{ln2}{\lambda}
,\label{subeq:1}
\end{equation}
here the decay constant $\mathcal{\lambda}$ can be obtained by
\begin{equation}
\
\lambda={\nu}P,
\label{subeq:1}
\end{equation}
where $\nu$ is the assault frequency related to oscillation frequency $\mathcal{\omega}$ \cite{PhysRevC.93.024612}. It can be obtained by
\begin{equation}
\
\nu=\frac{\omega}{2\pi}=\frac{2E_{\nu}}{h}
,\label{subeq:1}
\end{equation}
where $h$ is the Planck constant. The zero-point vibration energy $E_{\nu}$ is proportional to proton radioactivity energy $Q_p$ \cite{Poenaru1986}, which can be obtained by
\begin{equation}
\
E_{\nu}=\left\{\begin{array}{llll}

0.1045Q_p,&\text{for even-$Z$-even-$N$-parent nuclei},\\
0.0962Q_p,&\text{for odd-$Z$-even-$N$-parent nuclei},\\
0.0907Q_p,&\text{for even-$Z$-odd-$N$-parent nuclei},\\
0.0767Q_p,&\text{for odd-$Z$-odd-$N$-parent nuclei}.\\
\end{array}\right.
\label{subeq:1}
\end {equation}
$P$, the semiclassical WKB barrier penetration probability, can be calculated by
\begin{equation}
\
P=\exp(-2{\int_{r_{\text{in}}}^{r_{\text{out}}} k(r) dr})
,\label{subeq:1}
\end{equation}
where $k(r)=\sqrt{\frac{2\mu}{{\hbar}^2}|Q_p-V(r)|}$ is the wave number of the emitted proton. $\mu=\frac{{m_1}{m_2}}{{m_1}+{m_2}}$ denotes the reduced mass of the two-body system with $m_1$ being the daughter nucleus mass and $m_2$ being the proton mass. $\hbar$ is the reduced Planck constant. $r$ is the distance between the centers of the emitted proton and of the daughter nucleus. $V(r)$ and $Q_p$ are the total proton-core interaction potential and proton radioactivity energy, respectively. $r_{\text{in}}$ and $r_{\text{out}}$ are the classical turning points, they satisfy the conditions $V (r_{\text{in}}) = V (r_{\text{out}}) =Q_p$.

The total interaction potential $V(r)$, between the emitted proton and daughter nucleus, is composed of the nuclear potential $V_N(r)$, Coulomb potential
$V_C(r)$ and centrifugal potential $V_l(r)$. It can be expressed as
\begin{equation}
\
V(r)=V_N(r)+V_C(r)+V_l(r)
.\label{subeq:1}
\end {equation}
In this work we adopt the proximity potential formalism to calculate the emitted proton-daughter nucleus nuclear potential $V_N(r)$. The details are given in Section 2.2.

$V_C(r)$ is a Coulomb potential hypothesized as the potential of a uniformly charged sphere of radius $R$:
\begin{equation}
\
V_C(r)=\left\{\begin{array}{ll}

\frac{Z_1Z_2e^2}{2R}[3-(\frac{r}{R})^2],&r<R,\\

\frac{Z_1Z_2e^2}{r},&r>R,

\end{array}\right.
\label{subeq:1}
\end {equation}
where $R=R_1+R_2$. $R_1$ and $R_2$ denote the radii of daughter nucleus and emitted proton, respectively. Various expressions for $R_i \ (i=1, 2)$ within different proximity potential formalisms are given in Section 2.2. $Z_1$ and $Z_2$ are the proton number of daughter nucleus and emitted proton, respectively.

For the centrifugal potential $V_l(r)$, we adopt the Langer modified form, because $l(l+1){\to}(l+\frac{1}{2})^2$ is a necessary correction for one-dimensional problems \cite{1995JMP....36.5431M}. It can be expressed as
\begin{equation}
\
V_l(r)=\frac{{\hbar}^2(l+\frac{1}{2})^2}{2{\mu}r^2}
,\label{subeq:1}
\end {equation}
where $l$ is the angular momentum taken away by the emitted proton. The minimum angular momentum $l_{\text{min}}$ taken away by the emitted proton can be obtained by the conservation laws of spin and parity.

\subsection{The proximity potential formalism}
\label{2.2}
In the present work we select 28 versions of proximity potential formalisms to calculate the proton-daughter nucleus nuclear potential $V_N(r)$, which are: (i) Bass73 \cite{BASS1973139,BASS197445} and its revised versions Bass77 \cite{PhysRevLett.39.265} and Bass80 \cite{0954-3899-20-9-004}, (ii) CW76 \cite{CHRISTENSEN197619} and its revised versions BW91 \cite{0954-3899-20-9-004} and AW95 \cite{WINTHER1995203}, (iii) Ng$\hat{\text{o}}$80 \cite{NGO1980140}, (iv) Denisov \cite{DENISOV2002315} and its revised version Denisov DP \cite{PhysRevC.81.044615}, (v) Guo2013 \cite{GUO201354}, (vi) Prox.77 \cite{BLOCKI1977427} and its 12 modified forms on the basis of adjusting the surface energy coefficient \cite{MYERS19661,MOLLER1976502,PhysRevC.20.992,MOLLER1981117,0305-4616-10-8-011,MOLLER1988213,MOLLER1995185,PhysRevC.67.044316}, (vii) Prox.81 \cite{BLOCKI198153}, (viii) Prox.00 \cite{PhysRevC.62.044610} and its revised versions Prox.00 DP \cite{PhysRevC.81.044615}, Prox.2010 \cite{0256-307X-27-11-112402}, and Dutt2011 \cite{DUTT2011}. We find that only 7 of total 28 versions of the proximity potential formalisms can be used to deal with proton radioactivity, which are Bass73 \cite{BASS1973139,BASS197445}, CW76 \cite{CHRISTENSEN197619}, Denisov \cite{DENISOV2002315}, Denisov DP \cite{PhysRevC.81.044615}, Guo2013 \cite{GUO201354}, Prox.00 DP \cite{PhysRevC.81.044615} and Prox.2010 \cite{0256-307X-27-11-112402}. For the other versions of the proximity potential formalisms in process of handling some proton emitters, the classical turning points $r_{\text{in}}$ cannot be obtained through solving equation $V(r_{\text{in}})=Q_p$ due to the fact that the depth of $V(r)$ is located above the proton radioactivity energy $Q_p$. These 7 versions of proximity potential formalisms, which can be used to calculate the half-life of proton radioactivity, are expressed as follows:

\subsubsection{The proximity potential Bass73}
Based on the liquid drop model \cite{BASS1973139,BASS197445} Bass obtained the nuclear potential with the difference in surface energies between finite and infinite separation $\xi$. It is labeled as Bass73 and expressed as
\begin{equation}
\label{1}
\begin{aligned}
V_N(r)&=-4\pi\gamma\frac{dR_1R_2}{R}\exp(-\frac{\xi}{d})
\\&=\frac{-da_sA_1^{\frac{1}{3}}A_2^{\frac{1}{3}}}{R}\exp(-\frac{r-R}{d}),
\end{aligned}
\end{equation}
where $\gamma$ is the specific surface energy of the liquid drop model. $d=1.35$ fm is the range parameter. $a_s=17.0$ MeV represents the surface term in the liquid drop model mass formula. The sum of the half-maximum density radii $R=R_1+R_2=r_0(A_1^{\frac{1}{3}}+A_2^{\frac{1}{3}})$, where $r_0=1.07$ fm, $R_1$, $A_1$ and $R_2$, $A_2$ are the radii and mass numbers of daughter nucleus and emitted proton, respectively.

\subsubsection{The proximity potential CW76}
By analyzing the heavy-ion elastic scattering data, Christensen and Winther \cite{CHRISTENSEN197619} gave the empirical nuclear potential. It is labeled as CW76 and expressed as
\begin{equation}
\
V_N(r)=-50\frac{R_1R_2}{R_1+R_2}\phi(r-R_1-R_2)
,\label{subeq:1}
\end {equation}
with radius
\begin{equation}
\
R_i=1.233A_i^{\frac{1}{3}}-0.978A_i^{-\frac{1}{3}}{\ }(i=1,{\ }2)
.\label{subeq:1}
\end {equation}
The universal function $\phi(r-R_1-R_2)=\exp(-\frac{r-R_1-R_2}{0.63})$. 

\subsubsection{The proximity potential Denisov}
Denisov presented a expression for the nuclear potential part of ion-ion interaction potential by choosing 119 spherical or near spherical even-even nuclei around the $\beta$-stability line in the semi-microscopic approximation between all possible nucleus-nucleus combinations \cite{DENISOV2002315}. This proximity potential is labeled as Denisov and written as
\begin{equation}
\label{1}
\begin{aligned}
V_N(r)=&-1.989843\frac{R_1R_2}{R_1+R_2}\phi(\xi)\times[1+0.003525139\\
&\times(\frac{A_1}{A_2}+\frac{A_2}{A_1})^{\frac{3}{2}}-0.4113263(I_1+I_2)],
\end{aligned}
\end{equation}
here $I_1$ and $I_2$ are isospin asymmetry of daughter nucleus and emitted proton obtained by $I_i=\frac{N_i-Z_i}{N_i+Z_i}{\ }(i=1,{\ }2)$. $N_1$, $Z_1$, $R_1$ and $N_2$, $Z_2$, $R_2$ denote the neutron number, proton number and the effective nucleus radii of daughter nucleus and emitted proton, respectively. The effective nucleus radii can be obtained by
\begin{equation}
\
R_i=R_{ip}(1-\frac{3.413817}{R_{ip}^2})+1.284589(I_i-\frac{0.4A_i}{A_i+200}){\ }(i=1,{\ }2)
,\label{subeq:1}
\end {equation}
with proton radius $R_{ip}$ calculated by 
\begin{equation}
\
R_{ip}=1.240A_i^{\frac{1}{3}}(1+\frac{1.646}{A_i}-0.191I_i){\ }(i=1,{\ }2).
\label{subeq:1}
\end {equation}

The universal function $\phi(\xi)=\phi(r-R_1-R_2-2.65)$ is expressed as

%\ruleup
\begin{equation}
\
\phi(\xi)=\left\{\begin{array}{ll}
\{1-\xi^2[0.05410106\frac{R_1R_2}{R_1+R_2}\exp(-\frac{\xi}{1.760580})\\-0.5395420(I_1+I_2)\exp(-\frac{\xi}{2.424408})]\}\\\times\exp(\frac{-\xi}{0.7881663}),\ \xi\geq{0},\\
\\
1-\frac{\xi}{0.7881663}+1.229218\xi^2-0.2234277\xi^3\\-0.1038769\xi^4
-\frac{R_1R_2}{R_1+R_2}(0.1844935\xi^2\\+0.07570101\xi^3)+(I_1+I_2)(0.04470645\xi^2\\+0.03346870\xi^3),\ -5.65\leq\xi\leq0.\\
\end{array}\right.
\label{subeq:1}
\end {equation}
%\ruledown

\subsubsection{The proximity potential Denisov DP}

The proximity potential Denisov DP formalism is the modified version of Denisov which uses a more precise radius formula given by Ref. \cite{Royer2009} and expressed as
\begin{equation}
\
R_i=1.2332A_i^{\frac{1}{3}}+\frac{2.8961}{A_i^{\frac{2}{3}}}-0.18688A_i^{\frac{1}{3}}I_i{\ }(i=1,{\ }2)
.\label{subeq:1}
\end {equation}

\subsubsection{The proximity potential Guo2013}
By using the double folding model with density-dependent nucleon-nucleon interaction and fitting the universal functions of many reaction systems, Guo \textit{et al.} presented a new universal function of proximity potential model \cite{GUO201354}. This proximity potential is labeled as Guo2013 and expressed as
\begin{equation}
\
V_N(r)=4\pi{\gamma}b\frac{R_1R_2}{R_1+R_2}\phi(\xi)
,\label{subeq:1}
\end {equation}
where the diffuseness of nuclear surface $b$ is taken as unity, the surface coefficient $\gamma$ can be obtained by
\begin{equation}
\
\gamma=0.9517[1-1.7826(\frac{N-Z}{A})^2	]
,\label{subeq:1}
\end {equation}
here $N$, $Z$ and $A$ are neutron, proton and mass numbers, respectively, of the parent nucleus. $R_1$ and $R_2$ can be given by
\begin{equation}
\
\ R_i=1.28A_{i}^{\frac{1}{3}}-0.76+0.8A_{i}^{-\frac{1}{3}}{\ }(i=1,{\ }2)
.\label{subeq:1}
\end {equation}

The universal function $\phi(\xi)$ is expressed as
\begin{equation}
\
\phi(\xi)=\frac{p_1}{1+\exp({\frac{\xi+p_2}{p_3}})}
,\label{subeq:1}
\end {equation}
where $\xi=\frac{r-R_1-R_2}{b}$. $p_1$, $p_2$ and $p_3$ are adjustable parameters, whose values are -17.72, 1.30 and 0.854, respectively.

\subsubsection{The proximity potential Prox.00 DP}
Based on the proximity potential Prox.00 \cite{PhysRevC.62.044610}, Dutt and Puri presented a modified version of Prox.00 by choosing a more precise radius formula given by Royer and Rousseau \cite{Royer2009}. It is labeled as Prox.00 DP \cite{PhysRevC.81.044615} and expressed as
\begin{equation}
\
V_N(r)=4\pi{\gamma}b\bar{R}\phi(\xi)
,\label{subeq:1}
\end {equation}
where the mean curvature radius $\bar{R}$ can be obtained by
\begin{equation}
\
\bar{R}=\frac{C_1C_2}{C_1+C_2}
,\label{subeq:21}
\end {equation}
here $C_1$ and $C_2$ denote the matter radii of daughter nucleus and emitted proton, respectively. Based on the droplet model, they can be given by
\begin{equation}
\
C_i=c_i+\frac{N_i}{A_i}t_i{\ }(i=1,{\ }2)
,\label{subeq:22}
\end {equation}
where $c_1$ and $c_2$ are the charge distribution half-density radii of daughter nucleus and emitted proton. They can be expressed as
\begin{equation}
\
c_i=R_{i}(1-\frac{7b^2}{2R_{i}^2}-\frac{49b^4}{8R_{i}^4}){\ }(i=1,{\ }2)
,\label{subeq:23}
\end {equation}
with the nuclear charge radius \cite{Royer2009}
\begin{equation}
\
R_{i}=1.2332A_i^{\frac{1}{3}}+\frac{2.8961}{A_i^{\frac{2}{3}}}-0.18688A_i^{\frac{1}{3}}I_i{\ }(i=1,{\ }2)
.\label{subeq:24}
\end {equation}
Based on the droplet model \cite{Nerlo-Pomorska1994}, the neutron skin of daughter nucleus $t_1$ can be given by

\begin{equation}
t_1=\frac{3}{2}r_0\frac{SI_1-\frac{1}{12}gZ_1A_1^{-\frac{1}{3}}}{Q+\frac{9}{4}SA_1^{-\frac{1}{3}}}
,\label{subeq:25}
\end {equation}
where $r_0=1.14$ fm, the nuclear symmetric energy coefficient $S=32.65$ MeV, and $g=0.757895$ MeV. $Q=35.4$ MeV is the neutron skin stiffness coefficient. In this work $t_2=0$.

The surface energy coefficient $\gamma$ can be obtained by
\begin{equation}
\
\gamma=\frac{1}{4{\pi}r_0^2}(18.63-Q\frac{t_1^2}{2r_0^2})
.\label{subeq:1}
\end {equation}
The universal function $\phi(\xi)$ is expressed as
\begin{equation}
\
\phi(\xi)=\left\{\begin{array}{ll}

-0.1353+\sum_{n=0}^5\frac{c_n}{n+1}(2.5-\xi)^{n+1},&0<\xi\leq{2.5},\\

-0.09551\exp(\frac{2.75-\xi}{0.7176}),&\xi\geq{2.5},
\end{array}\right.
\label{subeq:1}
\end {equation}
here $\xi=\frac{r-C_1-C_2}{b}$ is the distance between the near surface of the daughter and emitted proton with the width parameter $b$ taken as unity. The values of $c_n$ are $c_0=-0.1886$, $c_1=-0.2628$, $c_2=-0.15216$, $c_3=-0.04562$, $c_4=0.069136$ and $c_5=-0.011454$ taken from Ref. \cite{PhysRevC.81.044615}.

\subsubsection{The proximity potential Prox.2010}
Using a suitable set of the surface energy coefficient $\gamma$, nuclear charge radius $R_i$ and universal function $\phi(\xi)$, Dutt and Bansal \cite{0256-307X-27-11-112402} presented another modified version of proximity potential Prox.00 \cite{PhysRevC.62.044610}. It is labeled as the proximity potential Prox.2010 and expressed as
\begin{equation}
\
V_N(r)=4\pi{\gamma}b\bar{R}\phi(\xi)
,\label{subeq:1}
\end {equation}
here the width parameter $b$ is taken as unity. The surface energy coefficient $\gamma$ is taken from Ref. \cite{MYERS19661} and expressed as
\begin{equation}
\
\gamma=\gamma_0[1-k_s(\frac{N-Z}{A})^2]
,\label{subeq:1}
\end {equation}
where the surface energy constant $\gamma_0=1.25284$ $\text{MeV}/\text{fm}^2$ and surface-asymmetry constant $k_s=2.345$ are given by Ref. \cite{PhysRevC.81.064608}.
The mean curvature radius $\bar{R}$, matter radius $C_i$, half-density radius of the charge distribution $c_i$, nuclear charge radius $R_{i}$ and neutron skin of daughter nucleus $t_1$ are the same as Eq. (\ref{subeq:21}), Eq. (\ref{subeq:22}), Eq. (\ref{subeq:23}), Eq. (\ref{subeq:24}) and Eq. (\ref{subeq:25}), respectively. Similarly, $t_2=0$.

The universal function $\phi(\xi)=\phi(\frac{r-C_1-C_2}{b})$ can be obtained by
\begin{equation}
\
\phi(\xi)=\left\{\begin{array}{lll}

-1.7817+0.9720\xi+0.143\xi^2-0.09\xi^3,\\\xi\leq0,\\
-1.7817+0.9720\xi+0.01696\xi^2-0.05148\xi^3,\\0\leq\xi\leq{1.9475},\\
-4.41\exp(\frac{-\xi}{0.7176}),{\ }\xi\geq{1.9475}.
\end{array}\right.
\label{subeq:1}
\end {equation}

\section{Results and discussion}

\begin{center}
\includegraphics[width=8.5cm]{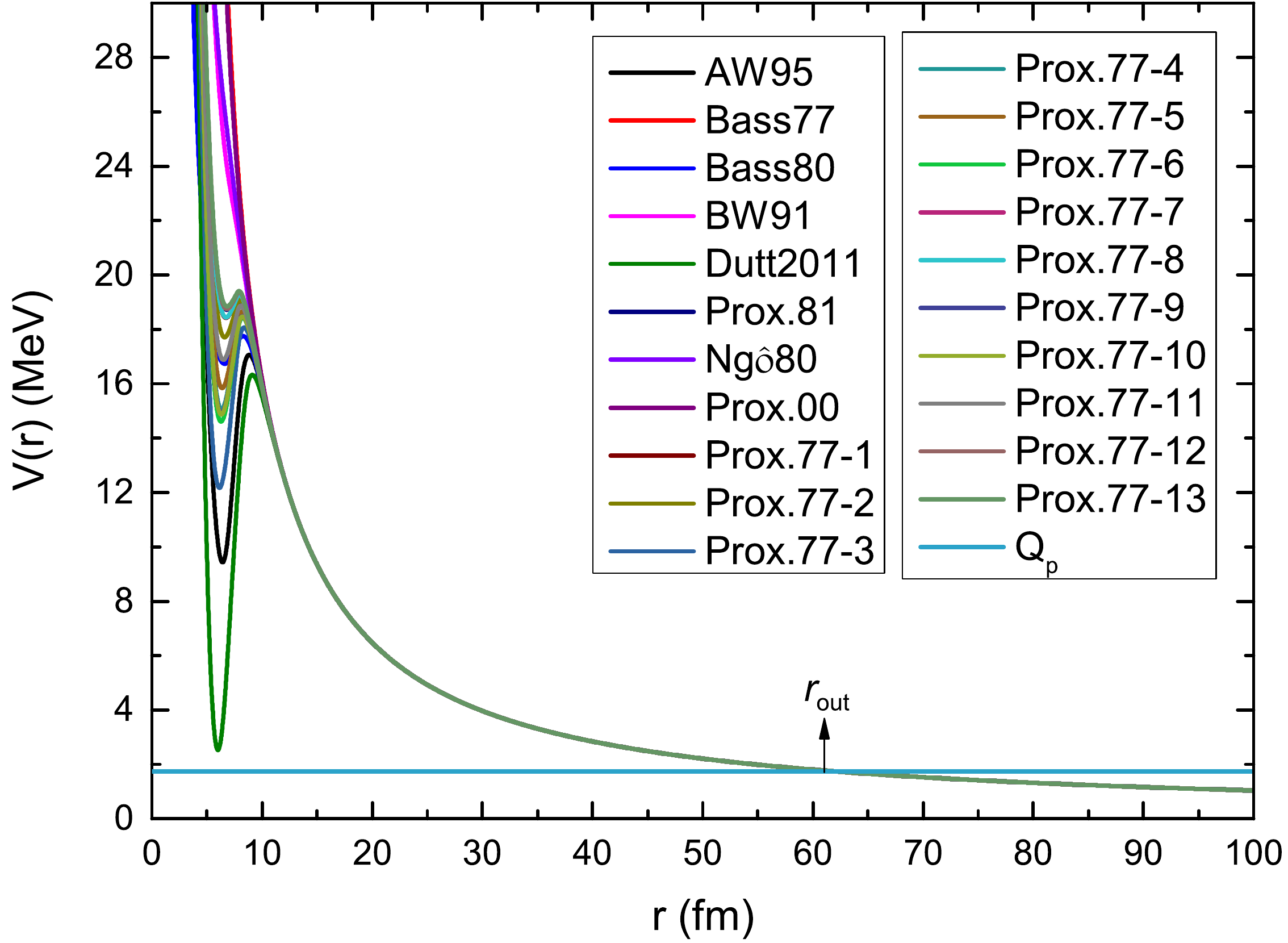}
\figcaption{\label{fig1}(Color online) The distributions of total emitted proton-core interaction potential $V(r)$ including the 21 versions of proximity potentials for $^{145}$Tm.}
\end{center}

\begin{center}
\includegraphics[width=8.5cm]{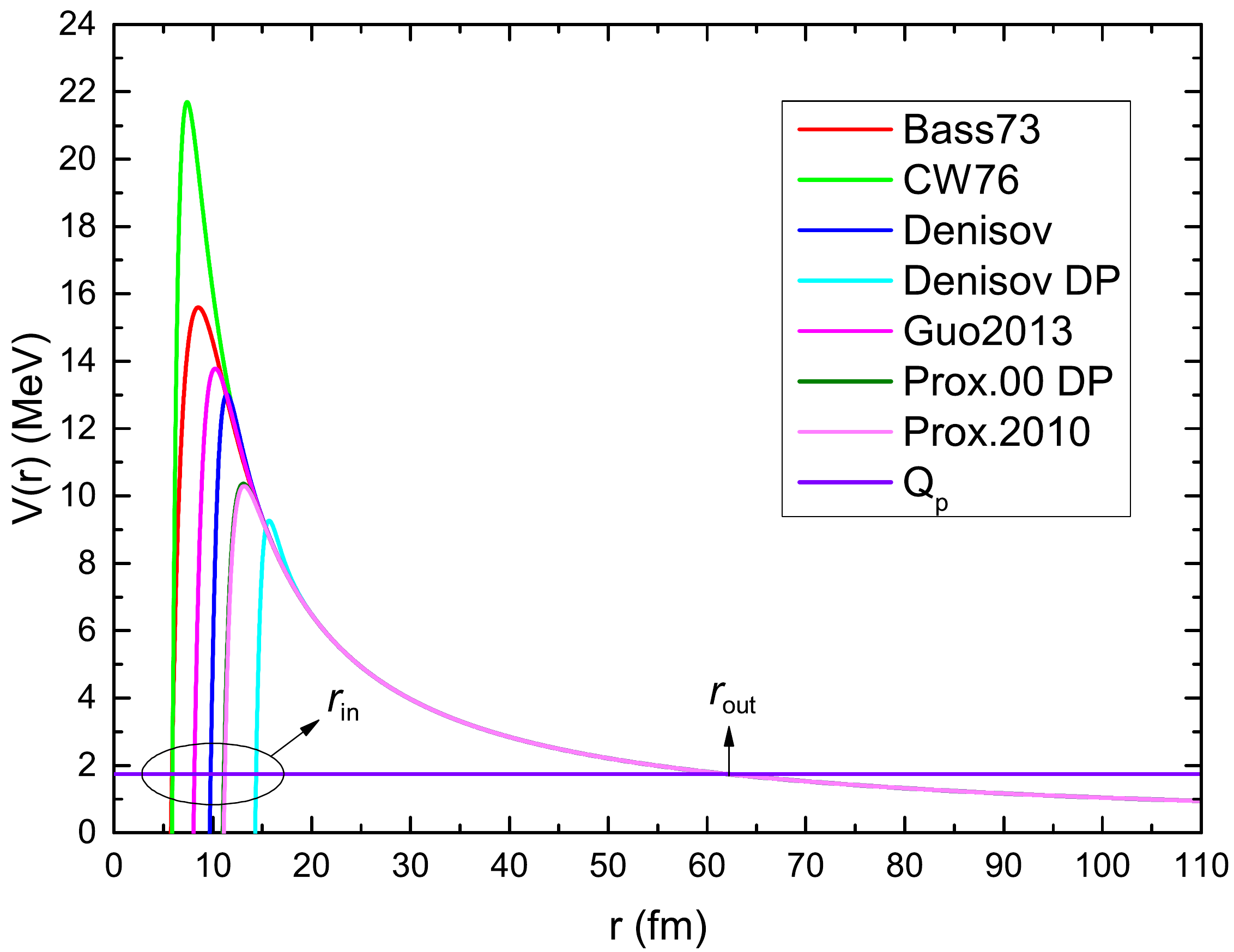}
\figcaption{\label{fig2}(Color online) The distributions of total emitted proton-core interaction potential $V(r)$ including the 7 versions of proximity potentials for $^{145}$Tm.}
\end{center}

The aim of the present work is to perform a comparative study of various proximity potentials when they are applied to the proton radioactivity of spherical proton emitters. We find that for these 21 versions of proximity potentials formalisms: Bass77 \cite{PhysRevLett.39.265}, Bass80 \cite{0954-3899-20-9-004}, BW91 \cite{0954-3899-20-9-004}, AW95 \cite{WINTHER1995203}, Ng$\hat{\text{o}}$80 \cite{NGO1980140}, Prox.77 \cite{BLOCKI1977427} and its 12 modified versions \cite{MYERS19661,MOLLER1976502,PhysRevC.20.992,MOLLER1981117,0305-4616-10-8-011,MOLLER1988213,MOLLER1995185,PhysRevC.67.044316}, Prox.81 \cite{BLOCKI198153}, Prox.00 \cite{PhysRevC.62.044610} and Dutt2011 \cite{DUTT2011} are not suitable to calculate the proton radioactivity half-lives. In those cases, the classical turning points $r_{\text{in}}$ cannot be obtained through solving equation $V(r_{\text{in}})=Q_p$ due to the fact that the depth of the total interaction potential $V(r)$ between the emitted proton and the daughter nucleus is above the proton radioactivity energy $Q_p$ for some proton emitters. Let us consider $^{145}$Tm as an example. Fig. \ref{fig1} shows the total interaction potential $V(r)$ obtained from those 21 proximity potentials mentioned above. From this figure, we can clearly see that all the depths of $V(r)$ are above the $Q_p$. It indicates that the aforementioned 21 versions of proximity potential formalisms are inappropriate for dealing with proton radioactivity. In the remainder of this paper, we will investigate the proton radioactivity half-lives by using the other 7 versions of proximity potential formalisms i.e., Bass73 \cite{BASS1973139,BASS197445}, CW76 \cite{CHRISTENSEN197619}, Denisov \cite{DENISOV2002315}, Denisov DP \cite{PhysRevC.81.044615}, Guo2013 \cite{GUO201354}, Prox.00 DP \cite{PhysRevC.81.044615} as well as Prox.2010 \cite{0256-307X-27-11-112402}, and give the most suitable one in aspect of calculating the proton radioactivity half-lives.

The experimental data and calculations are summarized in Tables  \ref{table1} and \ref{table2}. In these tables, the first and second columns denote the proton radioactivity transition from the parent nucleus to daughter nucleus and the experimental proton radioactivity energy. The third and fourth columns represent the spin-parity transition from initial state to final state and the minimum angular momentum taken away by the emitted proton. The fifth column is the experimental proton radioactivity half-life denoted as ${T_{1/2}^\text{exp}}$. The calculated proton radioactivity half-lives by using the proximity potentials Bass73, CW76, Denisov, Denisov DP are respectively listed in the last four columns of Table \ref{table1} and denoted as ${T_{1/2}^\text{Bass73}}$, ${T_{1/2}^\text{CW76}}$, ${T_{1/2}^\text{Denisov}}$, ${T_{1/2}^\text{Denisov DP}}$. The calculations by adopting the the proximity potentials Guo2013, Prox.00 DP, Prox.2010 are respectively given in the last three columns of Table \ref{table2} and denoted as ${T_{1/2}^\text{Guo2013}}$, ${T_{1/2}^\text{Prox.00DP}}$, ${T_{1/2}^\text{Prox.2010}}$. Experimental proton radioactivity half-lives, spin and parity are taken from the latest evaluated nuclear properties table NUBASE2016 \cite{1674-1137-41-3-030001}, the proton radioactivity energies are taken from the latest evaluated atomic mass table AME2016 \cite{1674-1137-41-3-030002,1674-1137-41-3-030003}. 

From Tables 1 and 2, we clearly see that various calculations predict very different half-lives, and that  ${T_{1/2}^\text{Bass73}}$, ${T_{1/2}^\text{CW76}}$, ${T_{1/2}^\text{Denisov DP}}$, ${T_{1/2}^\text{Prox.00 DP}}$ as well as ${T_{1/2}^\text{Prox.2010}}$ are quite different from the respective experimental values. Furthermore, we find that ${T_{1/2}^\text{Guo2013}}$ is the closest to ${T_{1/2}^\text{exp}}$, while ${T_{1/2}^\text{Denisov}}$ can in general reproduce the experimental values in magnitude. Based on our comparative analysis we can explore what particular feature of a given potential impacts these differences between various theoretical calculations, as well as differences between theory and experiment. From Section 2.1 we can find that proton radioactivity half-life can be obtained by the assault frequency $\nu$, which is dependent on proton radioactivity energy $Q_p$, and barrier penetration probability $P$, which is related to the total emitted proton-daughter nucleus interaction potential $V(r)$ and $Q_p$. However, for a given proton emitter, $Q_p$ and the centrifugal barrier are fixed. Therefore, these are various versions of the nuclear potential, $V_N(r)$, and of the Coulomb potential, $V_C(r)$, which cause the differences between calculated half-lives. In order to verity this conclusion, taking $^{145}$Tm for instance, we plot its 7 versions of total emitted proton--daughter nucleus interaction potential distributions in Fig. \ref{fig2}. From this figure, we can find that for various versions of $V(r)$ the values of $r_{\text{out}}$ are the same, while the values of $r_{\text{in}}$ are different. Moreover, the heights of $V(r)$ are different. The minimal value of $r_{\text{in}}$ and highest height of $V(r)$ in the CW76 formalism caused the minimum barrier penetration probability $P$. Thus ${T_{1/2}^\text{CW76}}$ is the maximum calculation in Table \ref{table1} and \ref{table2}. Similarly, the maximum value of $r_{\text{in}}$ and the lowest height of $V(r)$ in the Denisov DP formalism caused the smaller ${T_{1/2}^\text{Denisov DP}}$ than ${T_{1/2}^\text{exp}}$.

\begin{center}
\includegraphics[width=8.5cm]{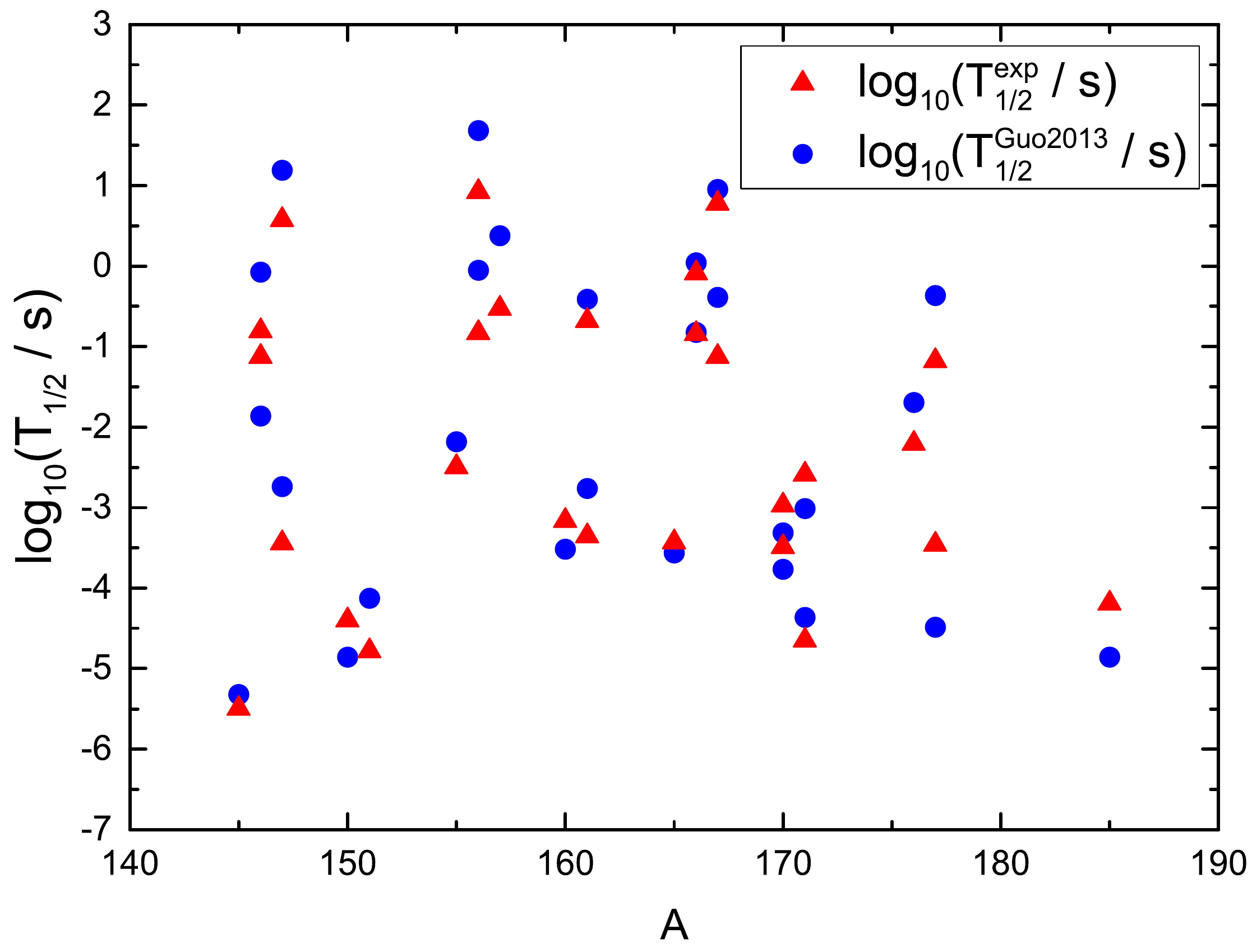}
\figcaption{\label{fig3}(Color online) Logarithmic half-lives of experimental and calculated data. The red triangles and blue circles denote the logarithms of experimental half-lives $T_{1/2}^\text{exp}$ and calculated results $T_{1/2}^\text{Guo2013}$, respectively.}
\end{center}

\begin{center}
\includegraphics[width=8.5cm]{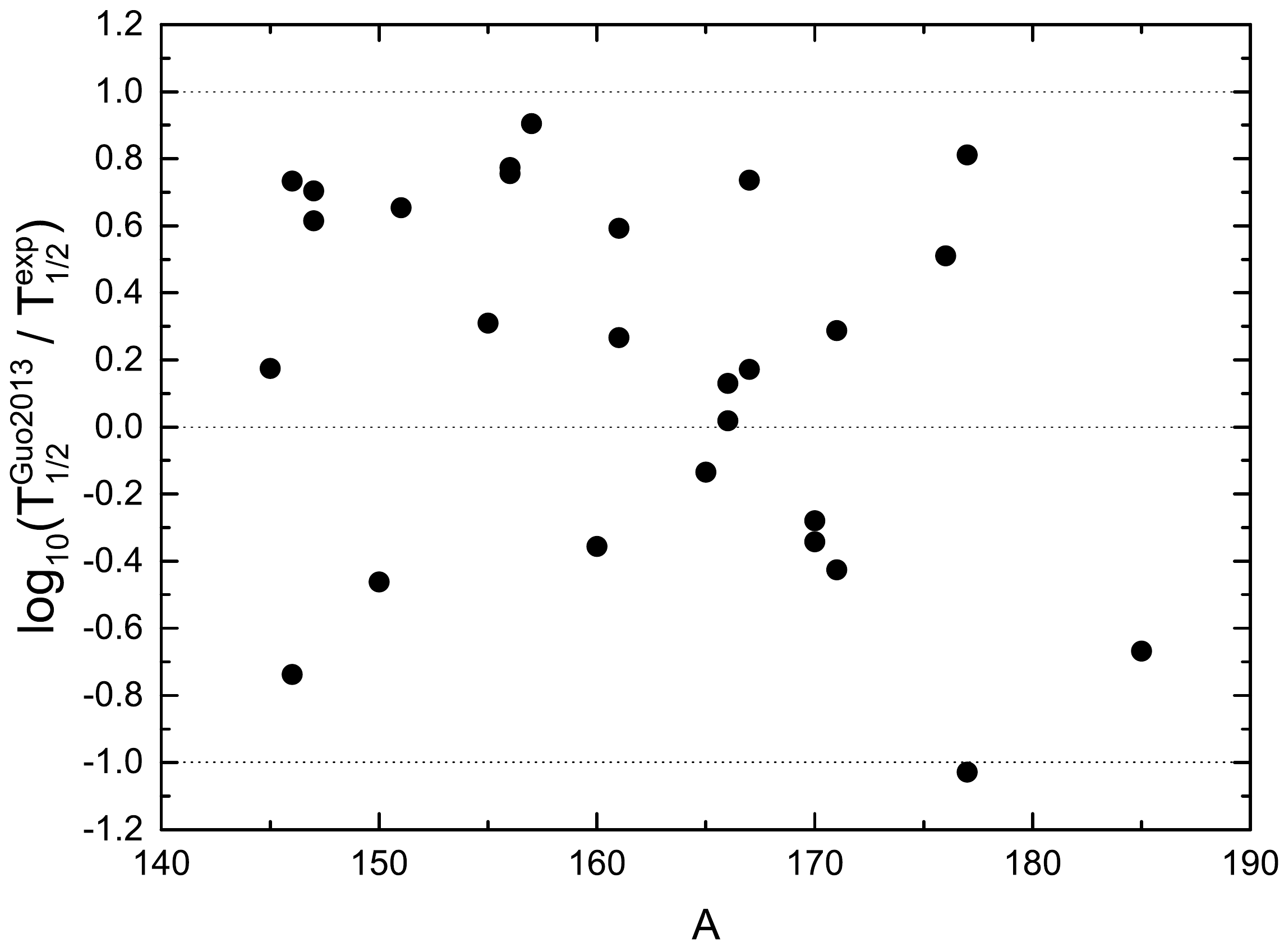}
\figcaption{\label{fig4}The logarithmic differences between $\log_{10}{T_{1/2}^{\text{Guo2013}}}$ and $\log_{10}{T_{1/2}^{\text{exp}}}$.}
\end{center}

The $T_{1/2}^\text{exp}$ and $T_{1/2}^\text{Guo2013}$ are plotted logarithmically in Fig. \ref{fig3}. From this figure, we can see that the ${T_{1/2}^\text{Guo2013}}$ can well reproduce the ${T_{1/2}^\text{exp}}$. In order to intuitively survey their deviations, we plot the difference between the logarithmic values of ${T_{1/2}^\text{Guo2013}}$ and ${T_{1/2}^\text{exp}}$ in Fig. \ref{fig4}. From this figure, we can clearly see that the values of $\log_{10}({T^\text{Guo2013}_{1/2}}/T^\text{exp}_{1/2})$ are mainly around zero, indicating that ${T_{1/2}^\text{Guo2013}}$ is in good agreement with the experimental data. In order to intuitively survey the deviations between proton radioactivity half-lives of calculations and experimental data, we calculate the standard deviation between the logarithmic values of calculations and experimental data $\sigma=\sqrt{\sum ({\log_{10}(T^{\rm{calc}}_{1/2}/T^{\rm{exp}}_{1/2}}))^2/n}$. The results for ${T_{1/2}^\text{Bass73}}$, ${T_{1/2}^\text{CW76}}$, ${T_{1/2}^\text{Denisov}}$, ${T_{1/2}^\text{Denisov DP}}$, ${T_{1/2}^\text{Guo2013}}$, ${T_{1/2}^\text{Prox.00 DP}}$ and ${T_{1/2}^\text{Prox.2010}}$ are listed in Table \ref{table3}. We find that the values of $\sigma$ for Bass73, CW76, Denisov DP, Prox.00 DP and Prox.2010 are large. This indicates that these five proximity potentials are not suitable to deal with the proton radioactivity of spherical proton emitters. For Denisov and Guo2013 the calculations can conform to the experimental data in magnitude. However, the value of $\sigma$ for Guo2013 is minimum. It demonstrates that the Guo2013 formalism can be adopted to obtain the most precise calculations of proton radioactivity half-lives of spherical proton emitters. In addition, Santhosh and Sukumaran studied the proton radioactivity half-lives using the CPPM and CPPMDN \cite{PhysRevC.96.034619}. In this work, we calculated the standard deviations between their calculations within two theoretical models and experimental data of spherical proton emitters. The standard deviations for CPPM is less than CPPMDN ($\sigma_{\text{CPPM}}=1.206$, $\sigma_{\text{CPPMDN}}=1.229$). This is consistent with their conclusion i.e., the calculated proton radioactivity half-lives by CPPM can better reproduce the experimental data than ones calculated by CPPMDN on the whole.

As an application, we use the proximity potential Guo2013 formalism to predict the proton radioactivity half-lives of 13 spherical proton emitters, whose proton radioactivity is energetically allowed or observed but not yet quantified in NUBASE2016 \cite{1674-1137-41-3-030001}. The predicated results are listed in Table \ref{table4}. In this table, the first four columns are same as Tables \ref{table1} and \ref{table2}. The fifth one is the predicted proton radioactivity half-life within proximity potential Guo2013 formalism also denoted as ${T_{1/2}^\text{Guo2013}}$. The spin and parity are taken from the NUBASE2016 \cite{1674-1137-41-3-030001}, the proton radioactivity energies are taken from the AME2016 \cite{1674-1137-41-3-030002,1674-1137-41-3-030003}. Based on the $\sigma=0.569$ of Guo2013 for 27 nuclei in the same region with predicted proton emitters, thus the predicted proton radioactivity half-lives are within a factor of 3.71.

Recent research has shown that the Geiger-Nuttall law can be used to describe the proton radioactivity isotopes with same angular momentum \cite{PhysRevC.96.034619}. In order to test our predictions, we plot the linear relationships between $\log_{10}{T_{1/2}^{\text{Guo2013}}}$ and $Q_p^{-1/2}$ of proton emitters $^{144,45,147}$Tm, $^{146}$Tm$^{\text{n}}$, $^{159,160,161}$Re and $^{164,165,166,167,169}$Ir$^{\text{m}}$ in Fig. \ref{fig5}. Each kind of isotopes have the same angular momentum $l$. In particular, the calculated proton radioactivity half-lives of $^{144}$Tm, $^{146}$Tm$^{\text{n}}$, $^{159}$Re, $^{164}$Ir$^{\text{m}}$ and $^{169}$Ir$^{\text{m}}$ are taken from Table \ref{table4}. From Fig. \ref{fig5}, we can find that the $\log_{10}{T_{1/2}^{\text{Guo2013}}}$ are linearly dependent on $Q_p^{-1/2}$. It is demonstrated that our predictions are credible.

\end{multicols}
\begin{center}
\tabcaption{Calculations of proton radioactivity half-lives using the proximity potential formalism Bass73, CW76, Denisov and Denisov DP. Elements with upper suffixes \lq{}m\rq{} and \lq{}n\rq{} indicate assignments to excited isomeric states (defined as higher states with half-lives greater than 100 ns). \lq{}()\rq{} means uncertain spin and/or parity. \lq{}\#\rq{} means values estimated from trends in neighboring nuclides with the same $Z$ and $N$ parities. The experimental proton radioactivity half-lives, spin and parity are taken from the NUBASE2016 \cite{1674-1137-41-3-030001}, the proton radioactivity energies are taken from the AME2016 \cite{1674-1137-41-3-030002,1674-1137-41-3-030003}. All proton radioactivity half-lives and energies are in units of \lq{}s\rq{} and \lq{}MeV\rq{}.}
\footnotesize
\label{table1}
\begin{longtable}{ccccccccc}
\hline Proton emission&$Q_p$&${J^{\pi}_{i}}\to{J^{\pi}_{f}}$&$l_{\text{min}}$&$T^\text{exp}_{1/2}$&${T_{1/2}^\text{Bass73}}$&${T_{1/2}^\text{CW76}}$&${T_{1/2}^\text{Denisov}}$&${T_{1/2}^\text{Denisov DP}}$\\&(MeV)&&&(s)&(s)&(s){}&(s)&(s)\\ \hline
\endfirsthead
\multicolumn{9}{c}%
{{Table 1. -- continued from previous page}} \\
\hline Proton emission&$Q_p$&${J^{\pi}_{i}}\to{J^{\pi}_{f}}$&$l_{\text{min}}$&$T^\text{exp}_{1/2}$&${T_{1/2}^\text{Bass73}}$&${T_{1/2}^\text{CW76}}$&${T_{1/2}^\text{Denisov}}$&${T_{1/2}^\text{Denisov DP}}$\\&(MeV)&&&(s)&(s)&(s){}&(s)&(s)\\ \hline
\endhead
\hline 
\endfoot
\hline
\endlastfoot
\noalign{\global\arrayrulewidth1pt}\noalign{\global\arrayrulewidth0.4pt} 
$^{145}$Tm$\to^{144}$Er$ $&1.741&${(11/2^-)}\to{0^+}$&5&$3.17\times10^{-6}$&$1.87\times10^{-4}$&$6.98\times10^{-4}$&$4.21\times10^{-7}$&$8.43\times10^{-10}$\\
$^{146}$Tm$\to^{145}$Er$ $&0.891&${(1^+)}\to{1/2^+\#}$&0&$1.55\times10^{-1}$&$4.69\times10^{0}$&$3.12\times10^{1}$&$1.87\times10^{-1}$&$1.47\times10^{-3}$\\
$^{146}$Tm$^\text{m}\to^{145}$Er$^\text{m}$&1.001&${(5^-)}\to{11/2^-\#}$&0&$7.50\times10^{-2}$&$7.53\times10^{-2}$&$5.04\times10^{-1}$&$3.11\times10^{-3}$&$2.55\times10^{-5}$\\
$^{147}$Tm$\to^{146}$Er$ $&1.059&${11/2^-}\to{0^+}$&5&$3.74\times10^{0}$&$6.50\times10^{2}$&$2.31\times10^{3}$&$1.26\times10^{0}$&$2.11\times10^{-3}$\\
$^{147}$Tm$^\text{m}\to^{146}$Er$ $&1.120&${3/2^+}\to{0^+}$&2&$3.60\times10^{-4}$&$1.43\times10^{-2}$&$9.06\times10^{-2}$&$3.36\times10^{-4}$&$2.07\times10^{-6}$\\
$^{150}$Lu$^\text{m}\to^{149}$Yb$ $&1.291&${(1^+,2^+)}\to{(1/2^+)}$&0&$4.00\times10^{-5}$&$7.69\times10^{-5}$&$5.09\times10^{-4}$&$3.09\times10^{-6}$&$2.67\times10^{-8}$\\
$^{151}$Lu$^\text{m}\to^{150}$Yb$ $&1.291&${(3/2^+)}\to{0^+}$&2&$1.65\times10^{-5}$&$6.01\times10^{-4}$&$3.74\times10^{-3}$&$1.34\times10^{-5}$&$8.31\times10^{-8}$\\
$^{155}$Ta$\to^{154}$Hf$ $&1.451&${(11/2^-)}\to{0^+}$&5&$3.20\times10^{-3}$&$2.75\times10^{-1}$&$9.53\times10^{-1}$&$5.05\times10^{-4}$&$9.12\times10^{-7}$\\
$^{156}$Ta$\to^{155}$Hf$ $&1.021&${(2^-)}\to{7/2^-\#}$&2&$1.49\times10^{-1}$&$7.81\times10^{0}$&$4.63\times10^{1}$&$1.44\times10^{-1}$&$7.82\times10^{-4}$\\
$^{156}$Ta$^\text{m}\to^{155}$Hf$ $&1.111&${(9^+)}\to{7/2^-\#}$&5&$8.39\times10^{0}$&$2.09\times10^{3}$&$7.05\times10^{3}$&$3.55\times10^{0}$&$5.86\times10^{-3}$\\
$^{157}$Ta$\to^{156}$Hf$ $&0.941&${1/2^+}\to{0^+}$&0&$2.96\times10^{-1}$&$1.47\times10^{1}$&$9.24\times10^{1}$&$4.72\times10^{-1}$&$3.47\times10^{-3}$\\
$^{160}$Re$\to^{159}$W $ $&1.271&${(4^-)}\to{7/2^-\#}$&0&$6.86\times10^{-4}$&$1.89\times10^{-3}$&$1.18\times10^{-2}$&$6.01\times10^{-5}$&$4.66\times10^{-7}$\\
$^{161}$Re$\to^{160}$W $ $&1.201&${1/2^+}\to{0^+}$&0&$4.40\times10^{-4}$&$1.09\times10^{-2}$&$6.75\times10^{-2}$&$3.37\times10^{-4}$&$2.57\times10^{-6}$\\
$^{161}$Re$^\text{m}\to^{160}$W $ $&1.321&${11/2^-}\to{0^+}$&5&$2.09\times10^{-1}$&$1.67\times10^{1}$&$5.56\times10^{1}$&$2.79\times10^{-2}$&$4.87\times10^{-5}$\\
$^{165}$Ir$^\text{m}\to^{164}$Os$ $&1.721&${(11/2^-)}\to{0^+}$&5&$3.72\times10^{-4}$&$1.15\times10^{-2}$&$3.84\times10^{-2}$&$1.97\times10^{-5}$&$3.79\times10^{-8}$\\
$^{166}$Ir$\to^{165}$Os$ $&1.161&${(2^-)}\to{(7/2^-)}$&2&$1.44\times10^{-1}$&$1.44\times10^{0}$&$8.08\times10^{0}$&$2.24\times10^{-2}$&$1.18\times10^{-4}$\\
$^{166}$Ir$^\text{m}\to^{165}$Os$ $&1.331&${(9^+)}\to{(7/2^-)}$&5&$8.12\times10^{-1}$&$4.80\times10^{1}$&$1.56\times10^{2}$&$7.53\times10^{-2}$&$1.31\times10^{-4}$\\
$^{167}$Ir$\to^{166}$Os$ $&1.071&${1/2^+}\to{0^+}$&0&$7.45\times10^{-2}$&$2.75\times10^{0}$&$1.64\times10^{1}$&$7.33\times10^{-2}$&$5.13\times10^{-4}$\\
$^{167}$Ir$^\text{m}\to^{166}$Os$ $&1.246&${11/2^-}\to{0^+}$&5&$6.00\times10^{0}$&$3.92\times10^{2}$&$1.27\times10^{3}$&$6.05\times10^{-1}$&$1.04\times10^{-3}$\\
$^{170}$Au$\to^{169}$Pt$ $&1.471&${(2^-)}\to{(7/2^-)}$&2&$3.25\times10^{-4}$&$1.65\times10^{-3}$&$9.17\times10^{-3}$&$2.54\times10^{-5}$&$1.41\times10^{-7}$\\
$^{170}$Au$^\text{m}\to^{169}$Pt$ $&1.751&${(9^+)}\to{(7/2^-)}$&5&$1.06\times10^{-3}$&$2.06\times10^{-2}$&$6.72\times10^{-2}$&$3.33\times10^{-5}$&$6.39\times10^{-8}$\\
$^{171}$Au$\to^{170}$Pt$ $&1.448&${(1/2^+)}\to{0^+}$&0&$2.23\times10^{-5}$&$2.93\times10^{-4}$&$1.75\times10^{-3}$&$7.85\times10^{-6}$&$5.95\times10^{-8}$\\
$^{171}$Au$^\text{m}\to^{170}$Pt$ $&1.702&${11/2^-}\to{0^+}$&5&$2.59\times10^{-3}$&$4.13\times10^{-2}$&$1.34\times10^{-1}$&$6.63\times10^{-5}$&$1.27\times10^{-7}$\\
$^{176}$Tl$\to^{175}$Hg$ $&1.261&${(3^-,4^-,5^-)}\to{(7/2^-)}$&0&$6.20\times10^{-3}$&$1.48\times10^{-1}$&$8.49\times10^{-1}$&$3.36\times10^{-3}$&$2.27\times10^{-5}$\\
$^{177}$Tl$\to^{176}$Hg$ $&1.155&${(1/2^+)}\to{0^+}$&0&$6.63\times10^{-2}$&$3.20\times10^{0}$&$1.82\times10^{1}$&$7.03\times10^{-2}$&$4.61\times10^{-4}$\\
$^{177}$Tl$^\text{m}\to^{176}$Hg$ $&1.962&${(11/2^-)}\to{0^+}$&5&$3.47\times10^{-4}$&$1.35\times10^{-3}$&$4.36\times10^{-3}$&$2.18\times10^{-6}$&$4.54\times10^{-9}$\\
$^{185}$Bi$^\text{m}\to^{184}$Pb$ $&1.607&${1/2^+}\to{0^+}$&0&$6.43\times10^{-5}$&$1.01\times10^{-4}$&$5.66\times10^{-4}$&$2.26\times10^{-6}$&$1.71\times10^{-8}$\\

\end{longtable}
\end{center}
\begin{multicols}{2}

\end{multicols}
\setcounter{table}{1}
\begin{center}
\tabcaption{Same as Table 1, but for calculations of proton radioactivity half-lives of spherical proton emitters using the proximity potential formalisms Guo2013, Prox.00 DP and Prox.2010.}
\footnotesize
\label{table2}
\begin{longtable}{cccccccc}
\hline Proton emission&$Q_p$ &${J^{\pi}_{i}}\to{J^{\pi}_{f}}$&$l_{\text{min}}$&$T^\text{exp}_{1/2}$&${T_{1/2}^\text{Guo2013}}$&${T_{1/2}^\text{Prox.00DP}}$&${T_{1/2}^\text{Prox.2010}}$\\&(MeV)&&&(s)&(s)&(s)&(s)\\ \hline
\endfirsthead
\multicolumn{8}{c}%
{{Table 2. -- continued from previous page}} \\
\hline Proton emission&$Q_p$ &${J^{\pi}_{i}}\to{J^{\pi}_{f}}$&$l_{\text{min}}$&$T^\text{exp}_{1/2}$&${T_{1/2}^\text{Guo2013}}$&${T_{1/2}^\text{Prox.00DP}}$&${T_{1/2}^\text{Prox.2010}}$\\&(MeV)&&&(s)&(s)&(s)&(s)\\ \hline
\endhead
\hline 
\endfoot
\hline
\endlastfoot
\noalign{\global\arrayrulewidth1pt}\noalign{\global\arrayrulewidth0.4pt} $^{145}$Tm$\to^{144}$Er$ $&1.741&${(11/2^-)}\to{0^+}$&5&$3.17\times10^{-6}$&$4.74\times10^{-6}$&$4.33\times10^{-8}$&$3.83\times10^{-8}$\\
$^{146}$Tm$\to^{145}$Er$ $&0.891&${(1^+)}\to{1/2^+\#}$&0&$1.55\times10^{-1}$&$8.39\times10^{-1}$&$2.96\times10^{-2}$&$2.73\times10^{-2}$\\
$^{146}$Tm$^\text{m}\to^{145}$Er$^\text{m}$&1.001&${(5^-)}\to{11/2^-\#}$&0&$7.50\times10^{-2}$&$1.37\times10^{-2}$&$4.96\times10^{-4}$&$4.57\times10^{-4}$\\
$^{147}$Tm$\to^{146}$Er$ $&1.059&${11/2^-}\to{0^+}$&5&$3.74\times10^{0}$&$1.54\times10^{1}$&$1.28\times10^{-1}$ &$1.13\times10^{-1}$\\
$^{147}$Tm$^\text{m}\to^{146}$Er$ $&1.120&${3/2^+}\to{0^+}$&2&$3.60\times10^{-4}$&$1.82\times10^{-3}$&$4.96\times10^{-5}$&$4.55\times10^{-5}$\\
$^{150}$Lu$^\text{m}\to^{149}$Yb$ $&1.291&${(1^+,2^+)}\to{(1/2^+)}$&0&$4.00\times10^{-5}$&$1.38\times10^{-5}$&$5.02\times10^{-7}$&$4.63\times10^{-7}$\\
$^{151}$Lu$^\text{m}\to^{150}$Yb$ $&1.291&${(3/2^+)}\to{0^+}$&2&$1.65\times10^{-5}$&$7.44\times10^{-5}$&$2.00\times10^{-6}$&$1.83\times10^{-6}$\\
$^{155}$Ta$\to^{154}$Hf$ $&1.451&${(11/2^-)}\to{0^+}$&5&$3.20\times10^{-3}$&$6.53\times10^{-3}$&$5.47\times10^{-5}$&$4.83\times10^{-5}$\\
$^{156}$Ta$\to^{155}$Hf$ $&1.021&${(2^-)}\to{7/2^-\#}$&2&$1.49\times10^{-1}$&$8.82\times10^{-1}$&$2.15\times10^{-2}$&$1.97\times10^{-2}$\\
$^{156}$Ta$^\text{m}\to^{155}$Hf$ $&1.111&${(9^+)}\to{7/2^-\#}$&5&$8.39\times10^{0}$&$4.79\times10^{1}$&$3.83\times10^{-1}$&$3.37\times10^{-1}$\\
$^{157}$Ta$\to^{156}$Hf$ $&0.941&${1/2^+}\to{0^+}$&0&$2.96\times10^{-1}$&$2.37\times10^{0}$&$7.71\times10^{-2}$&$7.11\times10^{-2}$\\
$^{160}$Re$\to^{159}$W $ $&1.271&${(4^-)}\to{7/2^-\#}$&0&$6.86\times10^{-4}$&$3.02\times10^{-4}$&$9.91\times10^{-6}$&$9.13\times10^{-6}$\\
$^{161}$Re$\to^{160}$W $ $&1.201&${1/2^+}\to{0^+}$&0&$4.40\times10^{-4}$&$1.72\times10^{-3}$&$5.60\times10^{-5}$&$5.16\times10^{-5}$\\
$^{161}$Re$^\text{m}\to^{160}$W $ $&1.321&${11/2^-}\to{0^+}$&5&$2.09\times10^{-1}$&$3.86\times10^{-1}$&$3.14\times10^{-3}$&$2.77\times10^{-3}$\\
$^{165}$Ir$^\text{m}\to^{164}$Os$ $&1.721&${(11/2^-)}\to{0^+}$&5&$3.72\times10^{-4}$&$2.72\times10^{-4}$&$2.30\times10^{-6}$&$2.03\times10^{-6}$\\
$^{166}$Ir$\to^{165}$Os$ $&1.161&${(2^-)}\to{(7/2^-)}$&2&$1.44\times10^{-1}$&$1.50\times10^{-1}$&$3.46\times10^{-3}$&$3.17\times10^{-3}$\\
$^{166}$Ir$^\text{m}\to^{165}$Os$ $&1.331&${(9^+)}\to{(7/2^-)}$&5&$8.12\times10^{-1}$&$1.10\times10^{0}$&$8.75\times10^{-3}$&$7.72\times10^{-3}$\\
$^{167}$Ir$\to^{166}$Os$ $&1.071&${1/2^+}\to{0^+}$&0&$7.45\times10^{-2}$&$4.05\times10^{-1}$&$1.23\times10^{-2}$&$1.13\times10^{-2}$\\
$^{167}$Ir$^\text{m}\to^{166}$Os$ $&1.246&${11/2^-}\to{0^+}$&5&$6.00\times10^{0}$&$8.94\times10^{0}$&$7.10\times10^{-2}$&$6.27\times10^{-2}$\\
$^{170}$Au$\to^{169}$Pt$ $&1.471&${(2^-)}\to{(7/2^-)}$&2&$3.25\times10^{-4}$&$1.71\times10^{-4}$&$3.99\times10^{-6}$&$3.65\times10^{-6}$\\
$^{170}$Au$^\text{m}\to^{169}$Pt$ $&1.751&${(9^+)}\to{(7/2^-)}$&5&$1.06\times10^{-3}$&$4.82\times10^{-4}$&$4.00\times10^{-6}$&$3.54\times10^{-6}$\\
$^{171}$Au$\to^{170}$Pt$ $&1.448&${(1/2^+)}\to{0^+}$&0&$2.23\times10^{-5}$&$4.33\times10^{-5}$&$1.35\times10^{-6}$&$1.24\times10^{-6}$\\
$^{171}$Au$^\text{m}\to^{170}$Pt$ $&1.702&${11/2^-}\to{0^+}$&5&$2.59\times10^{-3}$&$9.71\times10^{-4}$&$8.07\times10^{-6}$&$7.15\times10^{-6}$\\
$^{176}$Tl$\to^{175}$Hg$ $&1.261&${(3^-,4^-,5^-)}\to{(7/2^-)}$&0&$6.20\times10^{-3}$&$2.01\times10^{-2}$&$5.74\times10^{-4}$&$5.29\times10^{-4}$\\
$^{177}$Tl$\to^{176}$Hg$ $&1.155&${(1/2^+)}\to{0^+}$&0&$6.63\times10^{-2}$&$4.29\times10^{-1}$&$1.21\times10^{-2}$&$1.11\times10^{-2}$\\
$^{177}$Tl$^\text{m}\to^{176}$Hg$ $&1.962&${(11/2^-)}\to{0^+}$&5&$3.47\times10^{-4}$&$3.25\times10^{-5}$&$2.79\times10^{-7}$&$2.48\times10^{-7}$\\
$^{185}$Bi$^\text{m}\to^{184}$Pb$ $&1.607&${1/2^+}\to{0^+}$&0&$6.43\times10^{-5}$&$1.38\times10^{-5}$&$4.11\times10^{-7}$&$3.82\times10^{-7}$\\
\end{longtable}
\end{center}
\begin{multicols}{2}

%\begin{center}
%\includegraphics[width=8.5cm]{702.eps}
%\figcaption{\label{fig4}(Color online) Same as Fig. \ref{fig3}, but it is a local magnification of Fig. \ref{fig3}.}
%\end{center}

\end{multicols}
\begin{center}
\setcounter{table}{2}
\tabcaption{The standard deviations between logarithmic values of proton radioactivity half-lives of calculations and experimental data.}
\footnotesize
\label{table3}
\begin{longtable}{ccccccc}
\hline Bass73&CW76&Denisov&Denisov DP&Guo2013&Prox.00 DP&Prox.2010\\ \hline
\endfirsthead
\multicolumn{7}{c}%
{{Table 3. -- continued from previous page}} \\
\hline Bass73&CW76&Denisov&Denisov DP&Guo2013&Prox.00 DP&Prox.2010\\ \hline
\endhead
\hline \multicolumn{7}{r}{{Continued on next page}} \\
\endfoot
\hline
\endlastfoot
1.473&2.098&0.952&3.219&0.569&1.706&1.748\\
\end{longtable}
\end{center}
\begin{multicols}{2}

\end{multicols}
\begin{center}
\setcounter{table}{3}
\tabcaption{Same as Table 1, but for predicted radioactivity half-lives of spherical proton emitters, whose proton radioactivity is energetically allowed or observed but not yet quantified in NUBASE2016 \cite{1674-1137-41-3-030001}, within the proximity potential Guo2013 formalism.}
\footnotesize
\label{table4}
\begin{longtable}{ccccc}
\hline Proton emission&$Q_p$&${J^{\pi}_{i}}\to{J^{\pi}_{f}}$ &$l_{\text{min}}$&$T^\text{Guo2013}_{1/2}$\\&(MeV)&&&(s)\\ \hline
\endfirsthead
\multicolumn{5}{c}%
{{Table 4. -- continued from previous page}} \\
\hline Proton emission&$Q_p$&${J^{\pi}_{i}}\to{J^{\pi}_{f}}$ &$l_{\text{min}}$&$T^\text{Guo2013}_{1/2}$\\&(MeV)&&&(s)\\ \hline
\endhead
\hline \multicolumn{5}{r}{{Continued on next page}} \\
\endfoot
\hline
\endlastfoot
$^{144}$Tm$ \to^{143}$Er$ $&1.711&${(10^+)}\to{9/2^-\#}$&5&$9.03\times10^{-6}$\\
$^{146}$Tm$^\text{n}\to^{145}$Er$^\text{m}$&1.131&${(10^+)}\to{11/2^-\#}$&5&$2.03\times10^{0}$\\
$^{150}$Lu$ \to^{149}$Yb$ $&1.271&${(5^-,6^-)}\to{(1/2^+)}$&5&$1.59\times10^{-1}$\\
$^{151}$Lu$ \to^{150}$Yb$ $&1.241&${(11/2^-)}\to{0^+}$&5&$2.89\times10^{-1}$\\
$^{159}$Re$ \to^{158}$W $ $&1.591&${1/2^+\#}\to{0^+}$&0&$2.37\times10^{-7}$\\
$^{164}$Ir$ \to^{163}$Os$ $&1.561&${2^-\#}\to{7/2^-}$&2&$8.55\times10^{-6}$\\
$^{164}$Ir$^\text{m}\to^{163}$Os$ $&1.821&${(9^+)}\to{7/2^-}$&5&$6.09\times10^{-5}$\\
$^{165}$Ir$ \to^{164}$Os$ $&1.541&${1/2^+\#}\to{0^+}$&0&$1.89\times10^{-6}$\\
$^{169}$Ir$^\text{m}\to^{168}$Os$ $&0.765&${(11/2^-)}\to{0^+}$&5&$1.83\times10^{9}$\\
$^{169}$Au$ \to^{168}$Pt$ $&1.931&${1/2^+\#}\to{0^+}$&0&$7.16\times10^{-9}$\\
$^{172}$Au$ \to^{171}$Pt$ $&0.861&${(2^-)}\to{7/2^-}$&2&$7.60\times10^{4}$\\
$^{185}$Bi$^\text{n}\to^{184}$Pb$ $&1.703&${13/2^+\#}\to{0^+}$&6&$1.21\times10^{-1}$\\
$^{185}$Bi$ \to^{184}$Pb$ $&1.523&${9/2^-\#}\to{0^+}$&5&$2.47\times10^{-1}$\\
\end{longtable}
\end{center}
\begin{multicols}{2}

\begin{center}
\includegraphics[width=8.5cm]{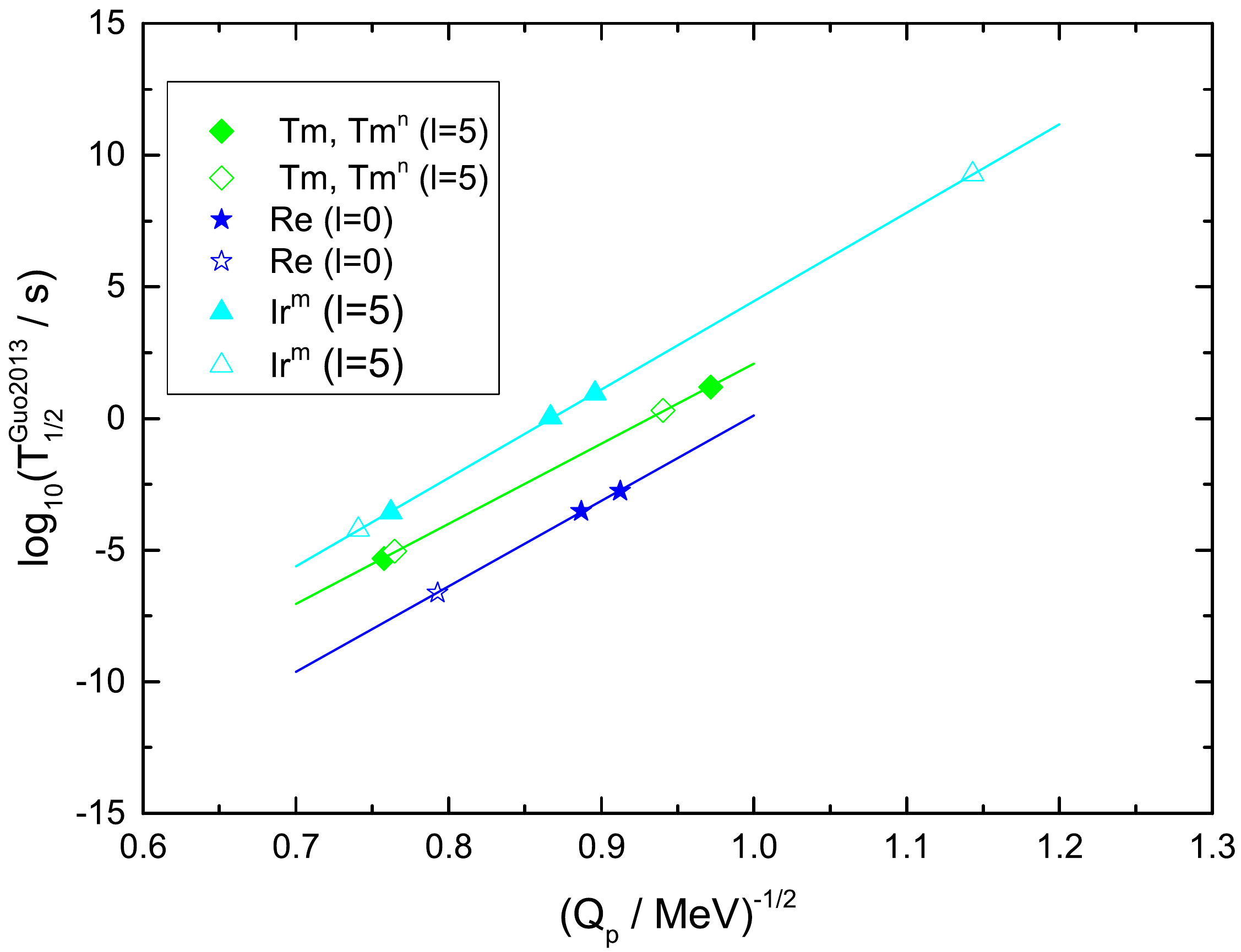}
\figcaption{\label{fig5}(Color online) The Geiger-Nuttall law for different cases of proton radioactivity between $\log_{10}{T_{1/2}^{\text{Guo2013}}}$ and $Q_p^{-1/2}$. The nuclei denoted as solid color and open symbols are taken from Table \ref{table2} and Table \ref{table4}, respectively.}
\end{center}
\section{Summary}
In summary, we performed a comparative study of 28 versions proximity potential formalisms applied to proton radioactivity half-lives of spherical proton emitters. The calculations show that the proximity potential Guo2013 formalism gives the lowest rms deviation in the description of the experimental half-lives of the known spherical protons emitters. As an application, we predict the proton radioactivity half-lives of 13 spherical proton emitters, whose proton radioactivity is energetically allowed or observed but not yet quantified in NUBASE2016, adopting the proximity potential Guo2013 formalism. Our calculated half-lives differ from the experiment values by a factor of 3.71 on average. This work will be used as a reference for experimental and theoretical research in the future.
\end{multicols}
\vspace{-1mm}
\centerline{\rule{80mm}{0.1pt}}
\vspace{2mm}

\begin{multicols}{2}

\end{multicols}

\clearpage
\end{CJK*}
\end{document}